\documentclass[twocolumn,a4paper,nofootinbib,
showpacs,aps,floatfix]{revtex4}
\usepackage{graphicx}
\usepackage{bm}
\usepackage{amsmath}
\def\la{\langle}
\def\ra{\rangle}
\def\ttot{\frac{\omega_T T_t}{2}}
\def\tfree{\frac{\omega_T T}{2}}

\begin {document}
% \title{Asymptotic behaviour in the Lamb-Dicke limit of residual Doppler effects
% in trapped ions}
\title{Motional frequency shifts of trapped ions in the Lamb-Dicke regime}
\author{I. Lizuain}
\email[Email address: ]{ion.lizuain@ehu.es}
\affiliation{Departamento de Qu\'\i mica-F\'\i sica,
Universidad del Pa\'\i s Vasco, Apdo. 644, Bilbao, Spain}
% \author{A. Ruschhaupt}
% \email[Email address: ]{a.ruschhaupt@tu-bs.de}
% \affiliation{Institut f\"ur Mathematische Physik, TU Braunschweig, Mendelssohnstrasse 3, 38106 Braunschweig, Germany} 
\author{J. G. Muga}
\email[Email address: ]{jg.muga@ehu.es} 
\affiliation{Departamento de Qu\'\i mica-F\'\i sica,
Universidad del Pa\'\i s Vasco, Apdo. 644, Bilbao, Spain} 
\author{J. Eschner}
\email[Email address: ]{juergen.eschner@icfo.es}
\affiliation{ICFO, Institut de Cienci\`es Fot\`oniques, 08860 Castelldefels, Barcelona, Spain}


%%%%%%%%%%%%%%%%%%%%%%%%%%%%%%%%%%%%%%%%%%%%%%%%%%%%%%%%%%%
\begin{abstract}
First order Doppler effects are usually ignored in laser driven trapped ions when the recoil frequency is 
much smaller than the trapping frequency (Lamb-Dicke regime).
This means that the central, carrier excitation band is supposed to be unaffected by vibronic transitions in which 
the vibrational number changes. While this is strictly true in the Lamb-Dicke limit (infinitely tight confinement),
the vibronic transitions do 
play a role in the Lamb-Dicke regime. In this paper we quantify the asymptotic behaviour of their effect with respect to the Lamb-Dicke parameter. In particular, we give analytical expressions for the frequency shift, ``pulling'' or ``pushing'', produced in
the carrier absorption band by the vibronic transitions both for Rabi and Ramsey 
schemes. This shift is shown to be independent of the initial vibrational state.    
\end{abstract}
\maketitle
%%%%%%%%%%%%%%%%%%%%%%%%%%%%%%%%%%%%%%%%%%%%%%%%%%%%%%%%%%%%
%%%%%%%%%%%%%%%%%%%%%%%%%%%%%%%%%%%%%%%%%%%%%%%%%%%%%%%%%%%%%%%%
\section{Introduction}
\begin{figure}[t]
\begin{center}
\includegraphics[height=6cm]{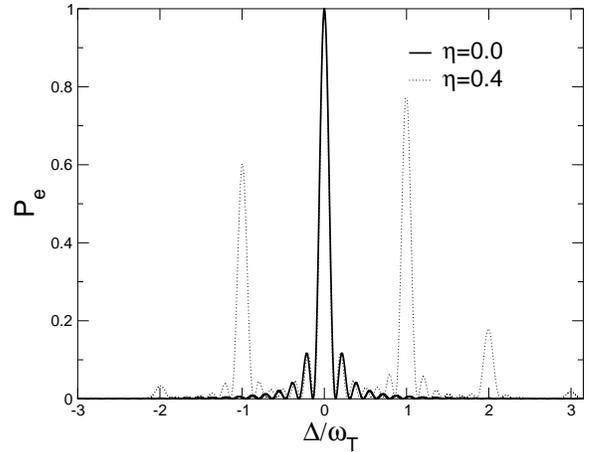}
\caption[]{Excited state probability of a trapped ion after a $\pi$-pulse has been applied. 
The ion is initially in the $|g,2\ra$ state.
% $\omega_T=(2\pi)\times 2.5$MHz, $\Omega_R=(2\pi)\times200$kHz. 
}
\label{eprob_rabi}
\end{center}
\end{figure}

There is currently much interest in laser cooled trapped ions because of metrological
applications as frequency standards, high precision spectroscopy, or the prospects of realizing quantum information processing \cite{LBMW03}. 
The absorption spectrum of a harmonically trapped (two-level) ion  
consists of a carrier band at the transition frequency 
$\omega_0$
and first order Doppler effect generated sidebands, equally spaced by the trap frequency $\omega_T$, see Fig. \ref{eprob_rabi}.
The excitation probability of a given sideband, and thus its intensity, depends critically on the so-called 
Lamb-Dicke (LD) parameter $\eta=\left[\hbar k_L^2/(2m\omega_T\right)]^{1/2}$, with $k_L$ being the 
driving laser wave number. If the LD regime is assumed ($\eta\ll1$),
the intensity of the $k$th red or blue sideband scales with $\eta^k$ \cite{LBMW03,WI79,WMILKM98}, $k=1,2,3,...$, so
the number of visible sidebands diminishes by decreasing $\eta$.
%
%
% is proportional to the modulus square of the 
% corresponding matrix element $\la n|e^{i\eta(a+a^\dag)}|n'\ra$, where 
% $\lbrace|n\ra\rbrace$ is the basis of the vibrational states
% and $\eta=\left(\hbar k_L^2/2m\omega_T\right)^{1/2}$ the so-called Lamb-Dicke (LD) paremeter.
% The intensities of these sidebands are proportional to
% $J_k^2(\eta)$, a function of the Lamb-Dicke parameter $\eta=k_Lx_0$ \cite{WI79}, see Fig. \ref{eprob_rabi}, where $x_0=\sqrt{\frac{\hbar}{2m\omega_T}}$ and $k_L$ is the laser wavenumber.
% As the $J_k^2(\eta)$ have negligible values when $k>\eta$, the number of 
% visible sidebands diminishes by decreasing $\eta$ 
% \cite{champenois04}.
%
It is then usually argued that in the LD regime
the absorption at the carrier frequency is free from first 
order Doppler effect \cite{WI79,D52,madej}. 
Of course this is only exact in the strict Lamb-Dicke limit, $\eta=0$, 
and for high precision spectroscopy, metrology, or quantum information applications, it is important to quantify  
the effect of the sideband transitions in the carrier peak, in other words,   
the asympotic behaviour, as $\eta\sim 0$, of the frequency shift of the carrier peak  
contaminated by vibronic, also called sideband, transitions in which the vibrational state changes.\footnote{Even though several transitions 
contribute to a given peak, it is named according to the dominant transition: thus we 
have a carrier peak 
or $k$-th sideband peaks.}
The inverse effect, in which the sideband is shifted by a non-resonant coupling to the carrier, has been previously studied in the field of trapped-ion based quantum computers \cite{HGRLBESB03,SHGRLEBB03}.
To get insight and the reference of analytical results, we shall examine a simplified one dimensional model neglecting decay from the excited state (resolved sideband regime
\cite{EMSB03}).  
The shift dependence on the various parameters (duration of the laser
pulses, Rabi frequency $\Omega_R$, $\omega_T$) will be explicitly obtained  
making use of a dressed state picture  
and a perturbation theory with respect to $\eta$.   
The cases of Rabi and Ramsey excitations will be examined separately since they may be
quite different quantitatively and have different applications as we shall see.   
%
%
%%%%%%%%%%%%%%%%%%%%%%%%%%%%%%%%%%%%%%%%%%%%%%%%%%%%%%%%%%%%%%%%
\subsection{Notation and Hamiltonian}
%
%%%%%%%%%%%%%%%%%%%%%%%% BEGIN FIGURE %%%%%%%%%%%%%%%%%%%%%%%%%%%%%%%%%%%%%%%%%%
%%%%%%%%%%%%%%%%%%%%%%%%% END FIGURE %%%%%%%%%%%%%%%%%%%%%%%%%%%%%%%%%%%%%%%%%%%
%
We consider a two level ion, with ground ($|g\ra$)
% $|g\ra=\left(\begin{array}{c} 1\\0\end{array}\right)$
and excited ($|e\ra$)
% $|e\ra=\left(\begin{array}{c} 0\\1\end{array}\right)$
states and transition frequency $\omega_0=\omega_e-\omega_g$, which is harmonically trapped and illuminated by 
a monochromatic laser of frequency $\omega_L$. 
In a frame rotating with the laser frequency, i.e., in a laser 
adapted interaction picture defined by $H_0=\frac{1}{2}\hbar\omega_L\sigma_z$, and in the usual (optical) Rotating Wave Approximation
(RWA), the ion is described by the time independent Hamiltonian \cite{CBZ94,LM06_b}
\begin{eqnarray}
H&=&\hbar\omega_T\left(a^\dag a+1/2\right)-\frac{\hbar\Delta}{2}\sigma_z
\nonumber
\\
&+&\frac{\hbar\Omega_R}{2}\left[
e^{i\eta(a+a^\dag)}\sigma_++h.c\right],
\label{time_ind_ham}
\end{eqnarray}
where $\Delta=\omega_L-\omega_0$ is the frequency difference between the laser and the internal transition (detuning), $\sigma_z=|e\ra\la e|-|g\ra\la g|$, $\sigma_+=|e\ra\la g|$, $\sigma_-=|g\ra\la e|$, and $a,a^\dagger$ are annihilation and creation operators 
for the vibrational quanta.

\begin{figure}[t]
\begin{center}
\vspace{1cm}
\includegraphics[height=6cm]{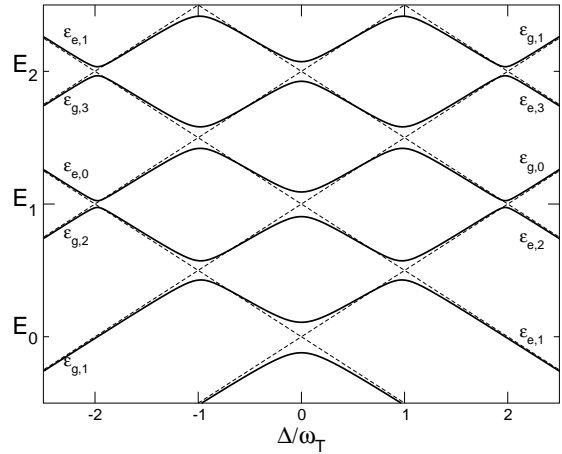}
\caption[]{Bare ($\Omega_R=0$, dashed line) and dressed ($\Omega_R/\omega_T=0.3$, solid line) energy levels
(in arbitrary units) as a function of the laser detuning. A not to small LD parameter $\eta=0.4$ has been 
intentionally chosen in order to highlight the higher order avoided crossings.}
\label{energy_levels}
\end{center}
\end{figure}

Let us denote by $|g,n\ra$ ($|e,n\ra$) the state of the ion in the ground (excited) internal state and in the $n^{th}$ motional level of the harmonic oscillator. In general 
the Hamiltonian (\ref{time_ind_ham}) will
couple internal and motional states.
The $\lbrace|g,n\ra,|e,n\ra\rbrace$ states form the ``bare'' basis of the system, i.e., the eigenstates of the bare hamiltonian $H_B=H(\Omega_R=0)$. 
The energy levels corresponding to the bare states are given by
\begin{eqnarray}
\epsilon_{g,n}&=&E_n+\frac{\hbar\Delta}{2},\nonumber\\
\epsilon_{e,n}&=&E_n-\frac{\hbar\Delta}{2},
\label{bare_levels}
\end{eqnarray}
with $E_n=\hbar\omega_T\left(n+1/2\right)$ being the energies of the harmonic oscillator.
These bare energy levels are plotted in Fig. \ref{energy_levels} (dotted lines) \cite{CBZ94,LM06_b} 
as a function of the detuning. 
They are degenerate when $\Delta=\pm k\omega_T$, $k=0,1,2,\hdots$, but the degeneracies 
are removed and become avoided crossings when the laser is turned on, see Fig. \ref{energy_levels} (solid lines).
At these avoided crossings transitions will occur between the involved (bare) states, 
which are  
nothing but the mentioned carrier ($k=0$) and sideband ($k\ge1$) transitions \cite{LBMW03}. The splitting
at each crossing gives the coupling strenght of a given transition \cite{LM06_b}, and the dynamics of the system
is then governed essentially by the reduced $2$-dimensional Hamiltonian of the involved levels.

Apart from these resonant transitions, off-resonant effects will also take place since, strictly speaking, the system is not
$2$-dimensional. In particular, near the atomic
transition resonance ($\Delta\sim0$), there will be a finite probability,
although small, of exciting higher order sidebands, which tends
to zero in the LD limit ($\eta\rightarrow0$). In this paper we study how these off-resonant effects behave within the LD regime,
when $\eta$ is made asymtotically small but not zero.
In particular, we study how these effects affect the excited (internal) state
probability, shifting the position of the central resonance,
which is crucial in fields such as atom interferometry \cite{oskay06} 
or atomic clocks with single trapped ions \cite{diddams_science_01}, where tiny deviations from the 
Doppler free form of the probability distribution could affect the accuracy of the measurements. Possible effects for state preparation in quantum information 
processing are also studied.

%
% %%%%%%%%%%%%%%%%%%%%%%%% BEGIN FIGURE %%%%%%%%%%%%%%%%%%%%%%%%%%%%%%%%%%%%%%%%%%
%%%%%%%%%%%%%%%%%%%%%%%%%% END FIGURE %%%%%%%%%%%%%%%%%%%%%%%%%%%%%%%%%%%%%%%%%%
%%%%%%%%%%%%%%%%%%%%%%%%%%%%%%%%%%%%%%%%%%%%%%%%%%%%%%%%%%%%%%%%%%%%%
\section{Frequency shift}
In precision spectroscopy experiments, the measured quantity is usually the excited 
(internal) state probability $P_e$, regardless of the vibrational quantum number $n$. 
If a general state of the trapped ion has the form
\begin{equation}
|\psi(t)\ra=\sum_{n=0}^{\infty}\left[g_n(t)|g,n\ra+e_n(t)|e,n\ra\right],
\label{eq3}
\end{equation}
the excited state probability $P_e$ will be given by
\begin{equation}
\label{prob_e}
P_e=tr\left(|\psi\ra\la\psi|e\ra\la e|\right)=\sum_{n=0}^\infty P_{e,n},
\end{equation}
where $P_{e,n}=\left|e_n(t)\right|^2$ is the probability of finding the $|e,n\ra$ state. In principle, the sum is over the infinite number 
of available vibrational quantum states, but it can be simplified if the LD regime is assumed.
In this regime the extension of the ion's wavefunction is much smaller than the driving 
laser wavelength, $\eta<<1$, and it is possible to expand the Hamiltonian 
(\ref{time_ind_ham}) in powers of $\eta$,
\begin{eqnarray}
\label{LD_ham}
H_{LD}&=&\hbar\omega_T\left(a^\dag a+1/2\right)-\frac{\hbar\Delta}{2}\sigma_z\nonumber\\
&+&\frac{\hbar\Omega_R}{2}\left[
(1+i\eta a+i\eta a^\dag)\sigma_++h.c\right],
\end{eqnarray}
which only couples, in first order, consecutive motional states. 
Then, if the ion is initially in the vibrational level $n_0$, only consecutive levels $n_0\pm 1$ will be coupled 
in a first order approximation. In other words, only carrier, first blue and first red sidebands will give appreciable
contributions to $P_e(\Delta\sim0)$.
It is thus possible to keep only the $n_0$ and $n_0\pm 1$ vibrational states and restrict our study
to the  $6$-dimensional subspace spanned by the $\lbrace|g,n_0\ra,|e,n_0\ra,|g,n_0\pm1\ra,|e,n_0\pm1\ra\rbrace$ 
bare states. 
The excited state probability (\ref{prob_e}) can then be approximated by
\begin{equation}
\label{pe_LD}
P_e\approx P_{e,n_0-1}+P_{e,n_0}+P_{e,n_0+1}
\end{equation}
in the LD regime. For all numerical cases examined, we have checked that adding further vibrational levels and using the Hamiltonian (1) 
leads to indistingishable results with respect to the six-state model if the LD condition is satisfied.

For an infinetely narrow trap
($\eta\rightarrow0$), only carrier transitions are driven (i.e., transitions in which the vibrational quantum 
number is not changed) and the central (carrier) peak of the excited state probability 
is exactly at atomic resonance, i.e., at $\Delta=0$. The generation
of blue and red sidebands will affect this distribution shifting the central maximum by $\delta$, where $\delta$ is the detuning that satisfies the maximum condition
\begin{equation}
\label{derivative}
\left.\frac{dP_e}{d\Delta}\right|_{\Delta=\delta}\approx\left.
\frac{d}{d\Delta}\left(P_{e,n_0-1}+P_{e,n_0}+P_{e,n_0+1}\right)\right|_{\Delta=\delta}=0
\end{equation}
and defines the ``frequency shift'' in the following sections.
This frequency shift can be understood as the error in determining the center of the resonance , i.e., the position 
of the maximum excitation. It will be shown that the position of this maximum, rather than coinciding with the
line center, varies periodically with the trap frequency $\omega_T$ when the sidebands are taken into account.

In the following sections this shift will be calculated in different excitation schemes, such as Rabi 
excitation (a single pulse, which is used in atomic clocks as well as quantum logic
applications); and Ramsey iterferometry 
(two pulses applied in atomic clocks and frequency standards).

%%%%%%%%%%%%%%%%%%%%%%%%%%%%%%%%%%%%%%%%%%%%%%%%%%%%%%%%%%%%%%%%%%%%%
\section{Single pulse (Rabi) excitation}
If an ion is prepared in $|\psi(t_i)\ra$ at an initial time $t_i$, the state of the system at a later time 
$t_f$ will be given by
\begin{eqnarray}
|\psi(t_f)\ra&=&e^{-iH(t_f-t_i)/\hbar}|\psi(t_i)\ra\nonumber\\
&=&\sum_\alpha e^{-i\epsilon_\alpha(t_f-t_i)/\hbar}|\epsilon_\alpha\ra\la\epsilon_\alpha|\psi(t_i)\ra,
\label{lc1}
\end{eqnarray}
where $|\epsilon_\alpha\ra$ ($\epsilon_\alpha$) are the $\alpha^{th}$ dressed states (energies) of the system, i.e., 
eigenstates (eigenenergies) of $H$.
We will consider first the case where a trapped ion is prepared in a given state $|g,n_0\ra$ at time $t_i=0$ 
and illuminated
by a single Rabi laser pulse for a time $\tau$.

The partial probabilities are easily obtained by
projecting the $|e,n\ra$ state on the state of the system at time $\tau$,
\begin{eqnarray}
P_{e,n}&=&\left|\la e,n|\psi(\tau)\ra\right|^2\nonumber\\
&=&\left|\sum_\alpha e^{-i\epsilon_\alpha\tau/\hbar}\la e,n|\epsilon_\alpha\ra\la\epsilon_\alpha|g,n_0\ra\right|^2,
\label{prob_en}
\end{eqnarray}
see an example in Fig. \ref{eprob_rabi}. For an infinitely narrow
trap ($\eta=0$),  $P_e(\Delta)$ is the well known Rabi pattern 
(solid line of Fig. \ref{eprob_rabi}). 
For non-zero LD parameters, sidebands are generated at integer multiples of the trap frequency $\omega_T$, (dotted
line in Fig. \ref{eprob_rabi}).
To obtain analytical expressions for these partial probabilities 
we shall follow the perturbative approach introduced in \cite{LM06_b}. 
%
%
%%%%%%%%%%%%%%%%%%%%%%%%%%%%%%%%%%%%%%%%%%%%%%%%%%%%%%%%%%%%%%%%%%%%%%%%%%%%
%
\subsection{Perturbative analysis: ``Semidressed'' states. }
\label{perturbation_section}
%
%A perturbative method will be followed in order to calculate the eigenergies and %eigenstates of $H_{LD}$. 
%The perturbative
%parameter should not be $\Omega_R$ as it is usually done, since the detuning can be 
%of the same order or larger. 
%
The perturbative approach in \cite{LM06_b} 
consists on dividing the Hamiltonian in Eq. (\ref{LD_ham}) as
\begin{equation}
H_{LD}=H_{SD}+V(\eta),
\end{equation}
with
\begin{eqnarray}
\label{sd_ham}
H_{SD}&=&\hbar\omega_T\left(a^\dag a+1/2\right)-\frac{\hbar\Delta}{2}\sigma_z
+\frac{\hbar\Omega_R}{2}\left(\sigma_++\sigma_-\right),
\nonumber\\
V(\eta)&=&\frac{\hbar\Omega_R\eta}{2}\left[i \left(a+ a^\dag\right)\sigma_++h.c\right],
\label{sd_coupling}
\end{eqnarray}
where $H_{SD}$ is a ``semi-dressed'' Hamiltonian, which describes the trapped ion coupled to a laser field,
but does not account for the coupling between different vibrational levels. This coupling is described by the
term $V(\eta)$. Note that $H_{LD}$ reduces to $H_{SD}$ in the LD limit ($\eta\rightarrow 0$), 
and $V(\eta)$ is a small perturbation of $H_{SD}$ in the LD regime, $\eta\ll1$.

Within this perturbative scheme, dressed states and energies of $H_{LD}$ are obtained up to leading 
order in the LD parameter $\eta$ in our $6$-dimensional sub-space, see Appendix \ref{perturbation}.
%%%%%%%%%%%%%%%%%%%%%%%%%%%%%%%%%%%%%%%%%%%%%%%%%%%%%%%%%%%%%%%%%%%%%%%%%%%%
%
\subsection{Excited state probability}
With the expressions of the dressed energies (\ref{dressed_energies}) and 
dressed states (\ref{dressed_states}) of $H_{LD}$,
one finds, after some lengthy algebra from Eq. (\ref{prob_en}), that the probability of finding the ion in the internal excited state after a laser pulse of duration
$\tau$ is given, for the three relevant motional levels, by 
\begin{eqnarray}
\label{prob_rabi_01}
P_{e,n_0-1}&=&n_0\frac{\eta^2\Omega_R^2}{\Omega^2\left(\omega_T^2-\Omega^2\right)^2}\nonumber\\
&\times&\left[\left(\Delta-\omega_T\right)\Omega
\cos\frac{\Omega \tau}{2}\sin\frac{\omega_T \tau}{2}\right.\nonumber\\
&+&\left.\left(\Omega^2-\Delta\omega_T\right)
\sin\frac{\Omega \tau}{2}\cos\frac{\omega_T \tau}{2}\right]^2,\nonumber\\
\label{prob_rabi_02}
P_{e,n_0}&=&\left(\frac{\Omega_R}{\Omega}\right)^2\sin^2\frac{\Omega t}{2}+
\frac{\eta^2\Omega_R^4}{4\Omega^2}(2n_0+1)\sin\frac{\Omega\tau}{2}\nonumber\\
&\times&\left[\frac{\sin\left(\omega_T\tau-\Omega\tau/2\right)}{(\omega_T-\Omega)^2}
-\frac{\sin\left(\omega_T\tau+\Omega\tau/2\right)}{(\omega_T+\Omega)^2}\right],\nonumber\\
\label{prob_rabi_03}
P_{e,n_0+1}&=&(n_0+1)\frac{\eta^2\Omega_R^2}{\Omega^2\left(\omega_T^2-\Omega^2\right)^2}\nonumber\\
&\times&\left[\left(\Delta+\omega_T\right)\Omega
\cos\frac{\Omega \tau}{2}\sin\frac{\omega_T \tau}{2}\right.\nonumber\\
&-&\left.\left(\Omega^2+\Delta\omega_T\right)
\sin\frac{\Omega \tau}{2}\cos\frac{\omega_T \tau}{2}\right]^2,
\end{eqnarray}
where $\Omega\equiv\sqrt{\Omega_R^2+\Delta^2}$ is the effective (detuning dependent) Rabi frequency.

These probabilities are different from the ones obtained if counter rotating terms in
Hamiltonian (\ref{LD_ham}) are neglected after appying a motional or vibrational RWA. In this case, instead
of a six-dimensinal model, three $2$-dimensional models are solved \cite{LBMW03}, 
to yield 
\begin{equation}
\label{VRWA_probs}
P_{e,n+k}=\left|\frac{\Omega_{n,n+k}}{f_n^k}\right|^2\sin^2\frac{f_n^k\tau}{2},
\end{equation}
where $\Omega_{n,n+k}=\Omega_R\la n|e^{i\eta(a+a^\dag)}|n+k\ra$ and 
\begin{equation}
f_n^k=\sqrt{(\Delta-k\omega_T)^2+\Omega_{n,n+k}^2}.  
\end{equation}
These simplified expressions for the excited state probabilities give quite different
frequency shifts as discussed later, and do not add to one exactly at one particular value 
of the detuning. 
%
%%%%%%%%%%%%%%%%%%%%%%%% BEGIN FIGURE %%%%%%%%%%%%%%%%%%%%%%%%%%%%%%%%%%%%%%%%%%
\begin{figure}[t]
\begin{center}
\includegraphics[height=6cm]{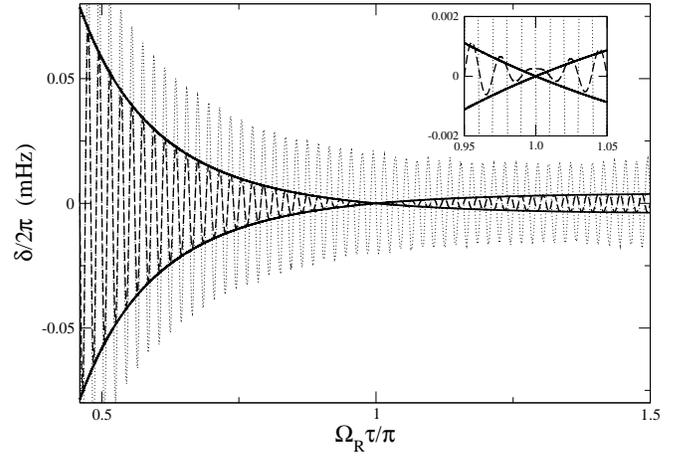}
\caption[]{Exact (numerical) (dashed line) frequency shift after a Rabi pulse of duration $\tau$ 
with a fixed Rabi frequency of $\Omega_R=2\pi\times 100$Hz and $\eta=0.05$.  
The solid lines represent approximate upper and lower bounds for the shift and the dotted one 
the shift obtained if the vibrational RWA is applied (using Eq.  (\ref{VRWA_probs}) for the probabilities).
Usual Paul (RF) traps have motional frequencies of a few MHz, but in this plot a trap frequency of 
$\omega_T=2\pi\times10$kHz has been considered in order to distinguish the fast oscilations. A zoom around 
the $\pi$-pulse is shown in the inset. }
\label{rabi_shift_fig}
\end{center}
\end{figure}
%%%%%%%%%%%%%%%%%%%%%%%%% END FIGURE %%%%%%%%%%%%%%%%%%%%%%%%%%%%%%%%%%%%%%%%%%%%
%
%
%
%
%
\subsection{Rabi frequency shift}
We are interested in the behaviour of $P_e$ near resonance, i.e., $\Delta\sim0$. If
only leading terms in $\Delta$ are kept and the maximum condition (\ref{derivative}) is applied to the
probabilities in Eq. (\ref{prob_rabi_03}),
it is found that for a weak laser (``weak'' meaning here that $\alpha\equiv\Omega_R/\omega_T\ll1$),
the frequency shift oscillates with the trap frequency $\omega_T$ as
%
%
% (a more general expression with $\alpha^2$ and $\alpha^3$ contributions
% \begin{eqnarray}
% \delta(\tau)&\approx&\Omega_R\eta^2\alpha^2f(\Omega_R\tau)\sin{\omega_T\tau}
% \times\nonumber\\
% &&\left(1+2\alpha\frac{\cos{\Omega_R\tau}\cos{\omega_T\tau}-1}{\sin{\Omega_R\tau}\sin{\omega_T\tau}}\right)
% \end{eqnarray}
%
\begin{eqnarray}
\label{rabi_shift}
\delta(\tau)&\approx&\Omega_R\eta^2\alpha^2f(\Omega_R\tau)\sin{\omega_T\tau},
\end{eqnarray}
with $f(\xi)$ being the function 
\begin{equation}
f(\xi)=\frac{\sin\xi}{\xi\sin \xi-4\sin^2 \frac{\xi}{2}}. 
\end{equation}

The exact (numerical) frequency shift $\delta$ is plotted in Fig. \ref{rabi_shift_fig} (dashed line)
as a function of the pulse duration time. 
The numerical calculations have been performed 
with the full Hamiltonian (\ref{time_ind_ham}), i. e., to all orders in the LD expansion, and 
with a large basis of bare states (more than 6). Here, and in all remaining figures, the numerical results
or the analytical approximation obtained in the LD regime are indistinguishable.

% Here, and in all remaining figures, the numerical
% calculations using the ``original Hamiltonian'' (\ref{time_ind_ham}) (to all orders in the LD expansion) with a large basis of bare states (more than 6), 
% or the analytical approximation obtained in the LD regime are indistinguishable.
The upper
and lower approximate bounds for the frequency shift (solid lines in Fig. \ref{rabi_shift_fig}) are obtained at the 
bounds of the fast oscillating term, i.e., replacing $\sin{\omega_T\tau}$ by $\pm1$ in Eq. (\ref{rabi_shift}),
\begin{equation}
\label{rabi_bounds}
\delta\left(\tau\right)\approx\Omega_R\eta^2\alpha^2f(\Omega_R\tau).
\end{equation}
If the applied pulse is a $\pi$-pulse ($\tau_\pi=\pi/\Omega_R$), the leading order contribution to the shift 
(\ref{rabi_shift}) vanishes and the next order in $\alpha$ has to be considered. 
Under the $\pi$-pulse condition there is some robustness against the shift error, reducing the frequency
shift to a pulling effect (i.e., a positive shift),
\begin{equation}
\label{rabi_shift_pi_pulse}
\delta(\tau_\pi)\approx\Omega_R\eta^2\alpha^3\cos^2{\frac{\omega_T\tau_\pi}{2}},
\end{equation}
which is not zero (see inset in 
Fig. \ref{rabi_shift_fig}) except for the values of $\Omega_R$ that make the 
argument of the cosine a multiple of $\pi/2$. 

Remarkably, the general frequency shift (\ref{rabi_shift}) is independent of the initial vibrational 
quantum number $n_0$. This follows from the fact that the probability for the first 
red sideband is proportional to the initial motional state $n_0$ while the first blue sideband is proportional
to $n_0+1$, see Eqs. (\ref{prob_rabi_03}). When the maximum condition
(\ref{derivative}) is applied, the $n_0$'s are cancelled.
Moreover, the result is identical to the shift when the  
ion is initially in the lowest vibrational
state. In this case, the frequency shift is just due to the first blue sideband (no red sidebands exist) but $n_0=0$. 
%, and a naive expectation could be that . This is
%not the case however, since the shift is independent of $n_0$. 
This particular case 
can be solved exactly in a $4$-state model, without a perturbative approach, giving the same results, see the Appendix \ref{4state_model}. 

Note also that if the vibrational RWA is applied and the simplified
expressions for the probabilities of the excited states (\ref{VRWA_probs}) are used to compute
the frequency shift, quite different results are obtained (dotted line in Fig. \ref{rabi_shift_fig}), with particularly high relative errors near the $\pi$-pulse condition.

In quantum information applications, the parameters $\alpha$ and $\Omega_R$ 
are usually higher than in frequency standards since the speed of the operations 
is of importance, so that the shift of the carrier peak may be much larger. 
We have collected some typical numerical values in Tables I and II. 

\begin{widetext}
%%%%%%%%%%%%%%%%%%%%%%%%%%%%%%%%%%%%%%%%%%%%%%%%%%%%%%%%%%%%%%%%%%%%
\begin{center}
\begin{table}
\begin{tabular}{cccccc}
\hline
Ion    &    $\omega_T/2\pi$    &    $\eta$   & $\Omega_R/2\pi$ (Hz)& $\delta/2\pi$ (Hz)   &   Reference\\ 
\hline
\hline
\\
$^{40}$Ca$^+$ ($729n$m)  &  $1$MHz& $0.095$   & $10-100$  & $10^{-12}-10^{-9}$ &  \cite{champenois04}\\
$^{199}$Hg$^+$ ($282n$m)   & few MHz & $0.035$  & $10-20$& $10^{-14}-10^{-13}$ &\cite{riis04,udem01}\\
$^{88}$Sr$^+$ ($674n$m)   & $2.5$MHz & $0.042$  & $250-500$  & $10^{-9}-10^{-8}$ &  \cite{LGRS04}
% $\tau=1s,100ms,1ms$
\end{tabular}
\caption[]{Rabi one pulse excitation (clocks and frequency standards):  
for $^{199}Hg^+$, $\eta$ and $\delta$ have beenn calculated with 
$\omega_T/2\pi=10$MHz.}
\label{rabi_clock_table}
\end{table}
\end{center}
%%%%%%%%%%%%%%%%%%%%%%%%%%%%%%%%%%%%%%%%%%%%%%%%%%%%%%%%%%%%%%%%%%%%

%%%%%%%%%%%%%%%%%%%%%%%%%%%%%%%%%%%%%%%%%%%%%%%%%%%%%%%%%%%%%%%%%%%%
\begin{center}
\begin{table}
\begin{tabular}{cccccc}
\hline
Ion    &    $\omega_T/2\pi$    &    $\eta$    &   $\Omega_R/2\pi$ (kHz)  &   $\delta/2\pi$ (Hz)   &   Reference\\ 
\hline
\hline
\\
Ba$^+$ ($650n$m)  &  $50$kHz& $0.26$     & $1.5-15$  & $10^{-1}-10^2$ &  \cite{CZ95}\\
$^{40}$Ca$^+$ ($729n$m)   & $2$MHz & $0.03$  & $5$  & $10^{-5}$ &  \cite{kaler03}
\end{tabular}
\caption[]{Rabi one pulse excitation: Quantum information and quantum logic.}
\label{rabi_logic_table}
\end{table}
\end{center}
%%%%%%%%%%%%%%%%%%%%%%%%%%%%%%%%%%%%%%%%%%%%%%%%%%%%%%%%%%%%%%%%%%%%
\end{widetext}

\subsection{Fidelity for a $\pi/2$-pulse}

The oscillations of the carrier peak shift with respect to $\omega_T\tau$,
Eq. (\ref{rabi_shift}),
may affect other observables as well. As an example we find    
similar oscillations in the context of quantum state preparation.  
When applying a resonant $\pi/2$-pulse to a trapped ion initially in the ground state 
the internal state obtained for $\eta=0$ is 
\begin{equation}
\label{ideal}
|\psi_{id}\ra=\frac{1}{\sqrt 2}\left(|g\ra+i|e\ra\right).
\end{equation}
The contamination due to the higher order sidebands for non-zero $\eta$ 
will make the real internal state differ from this ideal state. 
%This effect is clear if we look at the central (carrier) line of the absortiom spectrum of %the trapped ion. 
%The intensity of the carrier will not only be reduced by $1-n\eta^2/2$. The position of %this maximum 
%will oscillate around the center ($\Delta=0$) as $\delta(\tau)\propto \sin (\omega_T\tau)$ %as we have already shown,
%Eq. (\ref{rabi_shift}). 

\begin{figure}[t]
\begin{center}
\includegraphics[height=6cm]{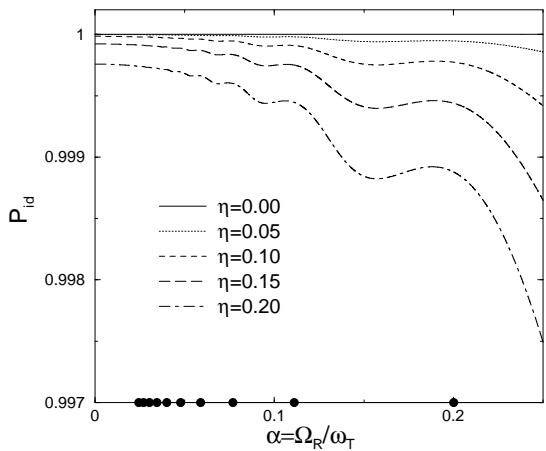}
\caption[]{Probability of detecting the ideal state $|\psi_{id}\ra$ as a 
function of $\alpha=\Omega_R/\omega_T$ for different LD parameters. 
The value of $\Omega_R\tau$ is fixed by the (resonant $\Delta=0$) $\pi/2$-pulse condition.
The circles
shown in the abscissa correspond to the maxima of the $\sin \frac{\pi}{2\alpha}$ function, i. e., 
$\alpha=\frac{1}{4n+1}$ with $n=0,1,2,\hdots$.}
\label{fidelity_fig}
\end{center}
\end{figure}

We now define the fidelity $\mathcal F$ as the probability of detecting the ideal state (\ref{ideal}), 
\begin{eqnarray}
\mathcal F=P_{id}
% &=&tr\left[\rho(\tau)|\psi_{id}\ra\la\psi_{id}|\right]\\
&=&tr\left[|\psi(\tau)\ra\la\psi(\tau)|\psi_{id}\ra\la\psi_{id}|\right]\\
&=&\frac{1}{2}\sum_{n=0}^\infty \left|g_n(\tau)-ie_n(\tau)\right|^2,
\end{eqnarray}
see Eq. (\ref{eq3}), 
where the sum is in principle over the infinite number of vibrational levels.
It is plotted in Fig. \ref{fidelity_fig} as a function of $\alpha=\Omega_R/\omega_T$. The fidelity is unity in 
the ``ideal" $\eta=0$ case but smaller otherwise. This fidelity oscillates also with the trap frequency, as it is observed in Fig. \ref{fidelity_fig}.
If a $\pi/2$-pulse is considered, we may rewrite the expression for the shift (\ref{rabi_shift}) as
\begin{eqnarray}
\delta\propto \sin{\omega_T\tau}= \sin \frac{\Omega_R\tau}{\alpha}=\sin\frac{\pi}{2\alpha}.
\end{eqnarray}
The maxima of the $\sin\frac{\pi}{2\alpha}$ 
function are marked with circles in the abscissa.  

%%%%%%%%%%%%%%%%%%%%%%%% BEGIN FIGURE %%%%%%%%%%%%%%%%%%%%%%%%%%%%%%%%%%%%%%%%%%
%%%%%%%%%%%%%%%%%%%%%%%%% END FIGURE %%%%%%%%%%%%%%%%%%%%%%%%%%%%%%%%%%%%%%%%%%%%
%
%
%
%
%
%
%
\section{Ramsey interferometry}
%
%
%
%
%
%
%
%% %%%%%%%%%%%%%%%%%%%%%%%% BEGIN FIGURE %%%%%%%%%%%%%%%%%%%%%%%%%%%%%%%%%%%%%%%%%%
\begin{figure}[t]
\includegraphics[height=6cm]{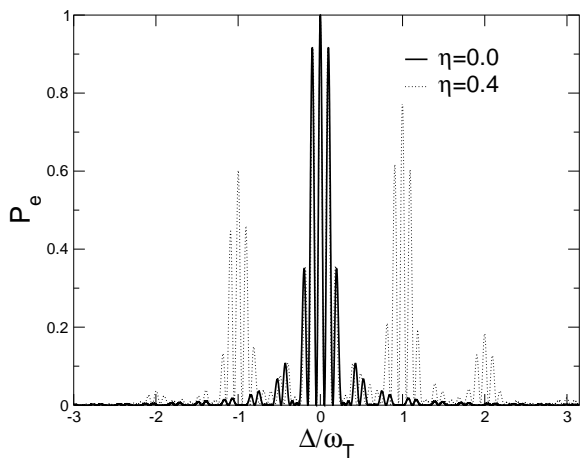}
\caption[]{Ramsey interference pattern for a non 
interaction time $T=2\tau$ and for different LD parameters.
% $\omega_T=2\pi\times 2.5$MHz, $\Omega_R=2\pi\times200$kHz. 
An ion initially in the $|g,2\ra$ state has been considered.}
\label{excited_prob_ramsey}
\end{figure}
\begin{figure}[t]
\begin{center}
% \includegraphics[height=6cm]{ramsey_shift_tau.eps}
% \vspace{1cm}
% \\
\includegraphics[height=6cm]{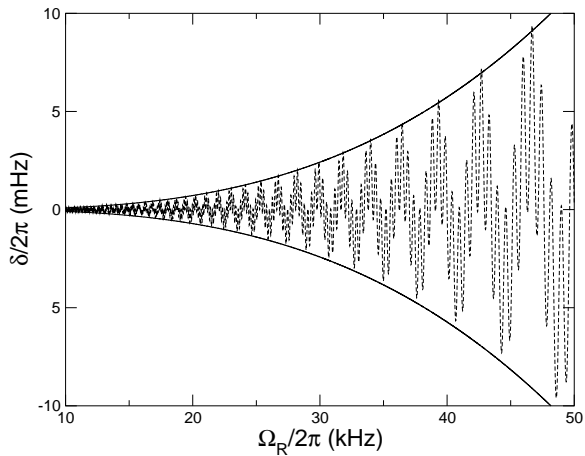}
% \vspace{1cm}
% \\
% \includegraphics[height=6cm]{ramsey_shift_10tau.eps}
 \caption[] {Exact frequency shift (dashed line) and approximate upper and lower bounds (solid lines)
as a function of the Rabi frequency
after a Ramsey $\pi/2$-pulse sequence with intermediate non-interaction time 
$T=5\tau$.
% in (a), $T=5\tau$ in (b) and $T=10\tau$ in (c).
An ion trapped within the LD regime ($\eta=0.04$) with a motional frequency $\omega_T=2\pi\times2$MHz has been considered.}
\label{ramsey_shift_fig}
\end{center}
\end{figure}

%%%%%%%%%%%%%%%%%%%%%%%%%% END FIGURE %%%%%%%%%%%%%%%%%%%%%%%%%%%%%%%%%%%%%%%%%% 
%
We may also calculate the frequency shift due to generation of higher order sidebands in a Ramsey scheme of 
two separated laser fields \cite{ramsey50}.
In these experiments with trapped ions, one ion prepared in the $|g,n_0\ra$ state is illuminated with two $\pi/2$-pulses ($\tau_{\pi/2}=\pi/2\Omega_R$) separated by a non-interaction or intermediate time $T$.
The state of the system at a time $2\tau_{\pi/2}+T$, after the two laser pulses,  
in the same laser-adapted interaction picture used before ($H_0=\frac{1}{2}\hbar\omega_L\sigma_z$)
is given by
\begin{equation}
\label{psit}
|\psi(2\tau_{\pi/2}+T)\ra=e^{-iH\tau_{\pi/2}/\hbar}e^{-iH_B T/\hbar}e^{-iH\tau_{\pi/2}/\hbar}|g,n_0\ra,
\end{equation}
where $H_B=H(\Omega_R=0)$ is the bare Hamiltonian governing the dynamics of the system in the intermediate region.
A simple generalization of Eq. (\ref{prob_en}) for two separated laser pulses, gives the probability 
for the different transitions,
\begin{eqnarray}
\label{ramsey_prob}
P_{e,n}&=&\bigg|\sum_\beta\sum_{j,k}\sum_\alpha e^{-i\epsilon_\beta\tau_{\pi/2}/\hbar}
e^{-i\epsilon_{j,k} T/\hbar}e^{-i\epsilon_{\alpha}\tau_{\pi/2}/\hbar}\nonumber\\
&&\la e,n|\epsilon_{\beta}\ra\la \epsilon_{\beta}
|j,k\ra\la j,k|
\epsilon_\alpha\ra\la\epsilon_\alpha|g,n_0\ra\bigg|^2,
\end{eqnarray}
with $\epsilon_{j,n}$ being the bare energies corresponding to the $|j,n\ra$ bare states ($j=g,e$), 
see Eq. (\ref{bare_levels}). The excited state probability distribution will
be given again by Eq. (\ref{prob_e}) in the general case, which is plotted in Fig. \ref{excited_prob_ramsey}.
It can be shown (Appendix  \ref{ramsey_shift_derivation}), that for weak lasers, the central maximum is shifted by
\begin{eqnarray}
\label{ramsey_shift}
\delta(T)&\approx&
% \Omega_R\eta^2\alpha^2\left(\frac{2}{2+\Omega_RT}\right)
% \cos\frac{\omega_T T_t}{2}\sin\frac{\omega_T T}{2}\\
\Omega_R\eta^2\alpha^2\left(\frac{2}{2+\Omega_RT}\right)\left[\cos{\ttot}\sin{\tfree}\right.\nonumber\\
&+&\left.\alpha\left(\cos^2{\ttot}+\sin^2{\tfree}\right)\right],
\end{eqnarray}
which is also independent of the initial vibrational quantum number $n_0$ and
where $T_t=2\tau_{\pi/2}+T$ is the total time of the experiment, see  Fig. \ref{ramsey_shift_fig}.
In the $T\rightarrow 0$ limit, this expression reduces to the one calculated for the Rabi 
case  when a $\pi$-pulse is applied, see Eq. (\ref{rabi_shift_pi_pulse}). 
For non-zero intermediate  times $T$, the leading order in $\alpha$ in Eq. (\ref{ramsey_shift})
may be written as
\begin{equation}
\delta(T)\approx
% \Omega_R\eta^2\alpha^2\left(\frac{2}{2+\Omega_RT}\right)\cos{\ttot}\sin{\tfree}
\frac{2\Omega_R\eta^2\alpha^2}{2+\Omega_RT}\cos{\ttot}\sin{\tfree},
\end{equation}
with approximate upper and lower bounds given by 
% (for $T_t\approx T$)
%
\begin{equation}
\label{ramsey_bounds}
\delta(T)\approx\pm \frac{2\Omega_R\eta^2\alpha^2}{2+\Omega_RT},
\end{equation}
see again Fig. \ref{ramsey_shift_fig} (solid lines). 
%%%%%%%%%%%%%%%%%%%%%%%% BEGIN FIGURE %%%%%%%%%%%%%%%%%%%%%%%%%%%%%%%%%%%%%%%%%%
% % %%Ramsey scheme figures
% %%%%%%%%%%%%%%%%%%%%%%%%% END FIGURE %%%%%%%%%%%%%%%%%%%%%%%%%%%%%%%%%%%%%%%%%%%

%
%
%%%%%%%%%%%%%%%%%%%%%%%%%%%%%%%%%%%%%%%%%%%%%%%%%%%%%%%%%%%%%%%%%%%%%%%%%%%%%%%%
\section{Discussion}
We have obtained analytical formulae that quantify the 
motional (sideband) effects in the carrier frequency peak of a trapped ion illuminated by a laser in the asymptotic Lamb-Dicke regime of tight confinement. 
Estimates of the importance of these effects for current or future experiments
have been provided in Tables \ref{rabi_clock_table} and \ref{rabi_logic_table}. 
The importance of the shift discussed here depends greatly on the application
and illumination scheme. Three different situations have been considered: 

(a) In single pulse Rabi interferometry, long laser pulses
are in principle desired in order to obtain narrow transitions, since the transition width is proportional 
to $1/\tau$, but this is limited by the stability of the laser and by the finite lifetime 
of the excited state. Typical laser pulses are of the order of miliseconds, that is,  
Rabi frequencies of tens to hundreds of Hertz if $\pi$-pulses (maximum excitation) are applied, 
which gives a frequency shift of $10^{-8}$ to $10^{-14}$ Hz, 
see Table \ref{rabi_clock_table}. Currently, the most accurate absolute measurement of an optical frequency
has fractional uncertainty of about $10^{-16}$, but 
frequency standards based on an optical transition in a single stored ion have the potential to reach
a fractional frequency uncertainty approaching $10^{-18}$ \cite{oskay06}. This means that the 
frequency shift found here corresponds to fractional errors of the order
of $10^{-24}-10^{-30}$ for typical optical transitions, which is far beyond the $10^{-18}$ level so that 
the shifts can be neglected in this context in the foreseeable future.

(b) This changes significantly for quantum information applications where 
fast operations are important and therefore the shifts 
are many orders of magnitude bigger even in the Rabi scheme, see Table \ref{rabi_logic_table}.

% Ions such as 
% $^{199}$Hg$^+$ at $282$nm \cite{diddams_science_01,oskay06} or $^{88}$Sr$^+$ at $674$nm \cite{margolis04,LGRS04} 
% are usually trapped in RF Paul traps with motional frequencies 
% $\omega_T/2\pi$ of a few Megahertz, leading to LD parameters around $5\times10^{-2}$, see also  \cite{G05}. 

(c) Back to metrology, the shift in the Ramsey scheme is more significant than in the Rabi 
scheme, 
because the illumination times 
are much shorter and thus the Rabi frequencies are
correspondingly higher.    
In recent Ramsey experiments with the $^{88}$Sr$^+$ ion at $674$nm a trap with motional frequency
$\omega_T\approx 2\pi\times2$MHz ($\eta\approx 0.042$) is driven by a laser with Rabi frequency $\Omega_R\approx 2\pi\times16$kHz, which
corresponds to laser pulses of several $\mu$s \cite{LGRS04}. 
Different intermediate times $T$ are used, ranging from $T=\tau$ to $T=10\tau$. It is clear from
Eq. (\ref{ramsey_bounds}) that the frequency shift decreases as the non interaction times $T$ increases.
With these data, Eq. (\ref{ramsey_bounds}) gives frequency shifts of 
$\delta\approx 2\pi\times 1m$Hz for $T=\tau$, which corresponds to a 
fractional error of order $10^{-18}$. The effect is therefore small today, but relevant 
for the most accurate experiments in the near future.  

Finally, a word is in order concerning the physical nature and interpretation 
of the shifts studied here. They are obviously associated with motional effects 
induced by the laser on the trapped ion, but they do not reflect energy level
shifts. 
Our frequency shifts are defined by the 
carrier peak displacement of the excitation probability. This probability is calculated with a linear combination of dressed states, 
as in Eqs. (\ref{prob_en}) and (\ref{ramsey_prob}). However, note that, while the eigenstates are affected (corrected) by the laser coupling of motional states 
characterized by the Lamb-Dicke parameter $\eta$, the energy eigenvalues remain unaffected
in first order in $\eta$, see the Appendix A.  
Indeed, the exact calculations of the shift 
(based on the general Hamiltonian (1) and converged with respect to the number of levels) are reproduced by  the approximations in which the eigenenergies remain unchanged, i.e.,
as in zeroth order with respect to $\eta$.  
The carrier peak shifts we have examined may in summary be viewed not as the 
result of energy-level shifts but due to dressed state corrections which affect the dynamics 
anyway. 
A consequence is their dependence on the illumination time.              

\section*{Acknowledgements}
This work has been supported by Ministerio de Edu\-ca\-ci\'on y Ciencia (FIS2006-10268-C03-01) and UPV-EHU (00039.310-15968/2004).

%
%
%APPENDIX
%\newpage
\appendix
\section{Perturbative corrections to the semidressed states.}
\label{perturbation}
The semidressed Hamiltonian (\ref{sd_ham}) is easily digonalized, with semidressed (i.e., zeroth order)
energies and states given by
\begin{eqnarray}
\epsilon_{n,\pm}^{(0)}&=&E_n\pm\frac{\hbar\Omega}{2},\\
|\epsilon_{n,\pm}^{(0)}\ra&=&\frac{1}{\sqrt{N_{\pm}}} \left( \frac{\Delta\pm\Omega}{\Omega_R}|g,n\ra+|e,n\ra\right),
\end{eqnarray}
$N_{\pm}$ being dimensionless normalization factors given by
\begin{equation}
\label{normalization}
N_{\pm}=\frac{\left(\Delta\pm\Omega\right)^2}{\Omega_R^2}+1=\frac{2\Omega}{\Omega_R^2}\left(\Omega\pm\Delta\right). 
\end{equation}

The dressed energies and states will be calculated by standard (time-independent) perturbation theory, with the
perturbation given by the coupling term $V(\eta)$, see Eq. (\ref{sd_coupling}). 
The matrix elements connecting the semidressed states are given in the LD regime by \cite{LM06_b}
\begin{eqnarray}
\label{matrix_elements}
&&\la \epsilon_{n,s}^{(0)}|V(\eta)|\epsilon_{n',s'}^{(0)}\ra
\nonumber\\
&=&i\eta s'\frac{\hbar\Omega}{\sqrt{N_sN_{s'}}}\left(\sqrt{n'}\delta_{n,n'-1}+
\sqrt{n'+1}\delta_{n,n'+1}\right)\left(1-\delta_{ss'}\right),\nonumber
\end{eqnarray}
where $s$ and $s'$ is a shorthand notation representing the sign ($s,s,'=\pm$).
Perturbation theory provides expressions for the dressed energies 
\begin{equation}
\label{dressed_energies}
\epsilon_{n,\pm}=\epsilon_{n,\pm}^{(0)}+{\cal{O}}(\eta^2)
\end{equation}
with no linear corrections, since the diagonal terms of the matrix elements (\ref{matrix_elements}) are zero. 
The dressed states (up to linear terms in $\eta$) are given by
\begin{eqnarray}
\label{dressed_states}
|\epsilon_{n-1,\pm}\ra&=&|\epsilon_{n-1,\pm}^{(0)}\ra
\pm\frac{i\eta\Omega_R\sqrt{n}}{-\omega_T\pm\Omega}|\epsilon_{n,\mp}^{(0)}\ra,\\
% +{\cal{O}}(\eta^2)
%
|\epsilon_{n,\pm}\ra&=&|\epsilon_{n,\pm}^{(0)}\ra
\pm\frac{i\eta\Omega_R\sqrt{n}}{\omega_T\pm\Omega}|\epsilon_{n-1,\mp}^{(0)}\ra\nonumber\\
&\pm&\frac{i\eta\Omega_R\sqrt{n+1}}{-\omega_T\pm\Omega}|\epsilon_{n+1,\mp}^{(0)}\ra,\\
% {\cal{O}}(\eta^2)\\
|\epsilon_{n+1,\pm}\ra&=&|\epsilon_{n+1,\pm}^{(0)}\ra
\pm\frac{i\eta\Omega_R\sqrt{n+1}}{\omega_T\pm\Omega}|\epsilon_{n,\mp}^{(0)}\ra.
% +{\cal{O}}(\eta^2)
\end{eqnarray}

%%%%%%%%%%%%%%%%%%%%%%%%%%%%%%%%%%%%%%%%%%%%%%%%%%%%%%%%%%%%%%%%%%
\section{$4$-state model, exact solution}
\label{4state_model}
If the ion is previously cooled down to its ground $|g,0\ra$ state (e. g., via sideband cooling), the 
problem becomes $4$-dimensional, since no red sideband will be involved, and analytical dressed states
can be obtained without using the perturbative treatment of Section \ref{perturbation_section}.
In this $4$-state model the excited state probability will then read
\begin{equation}
P_e\approx P_{e,0}+P_{e,1}
\end{equation}
within the LD regime. 
%This approximation will enable us to obtain analytical
%expressions of the dressed states without loosing any accuracy if a $|g,0\ra$ initial state %is considered, taking only
%into account $n=0$ and $n=1$ lowest motional levels.
The expressions for the dressed eigenergies are given by
\begin{eqnarray}
\epsilon_{s,s'}&=&\hbar\omega_T+s\frac{\hbar\nu_{s'}}{2},
\end{eqnarray}
where  $s$ and $s'$ is a shorthand notation representing a sign ($s,s'=\pm$).
The (angular) frequencies $\nu_\pm$ are defined
by $\nu_{\pm}\equiv\sqrt{(\omega_T\pm\Omega)^2+\eta^2\Omega_R^2}$, with 
$\Omega\equiv\sqrt{\Omega_R^2+\Delta^2}$ as usual. 
Near $\Delta=0$, these $\nu_\pm$ are frequencies shifted to the blue and red with respect to the trap frequency $\omega_T$; they correspond to transitions among the dressed levels and play an important role in the carrier frequency shift as we shall see.
The corresponding dressed eigenstates can be written as a function of the bare states,
\begin{eqnarray}
|\epsilon_{s,s'}\ra&=&\frac{1}{\sqrt{N_{s,s'}}}\left[
\frac{i}{\Omega_R}\left(\omega_T+s'\Omega-s\nu_{s'}\right)|g,0\ra\right.\nonumber\\
&+&\left.\frac{\eta\Omega_R}{s'\Omega-\Delta}|g,1\ra\right.\nonumber\\
&-&\left.\frac{i}{s'\Omega-\Delta}\left(\omega_T+s'\Omega-s\nu_{s'}\right)|e,0\ra\right.\nonumber\\
&+&\left.\eta|e,1\ra\right],
\end{eqnarray}
with $N_{ss'}$ being normalization factors. (Strictly  speaking, these states are ``partially''
dressed states in the sense that they are eigenstates of a part of the full Hamiltonian.) 

%From Eq. (\ref{prob_en}) and with an initial condition $|\psi(0)\ra=|g,0\ra$ the %probability 
%of finding the system in a $|e,n\ra$ will be given by,

If the ion is assumed initially in the ground $|g,0\ra$ state and is illuminated by a single laser pulse for a time $\tau$, 
the probability  of $|e,n\ra$ is 
\begin{eqnarray}
P_{e,n}&=&\left|\sum_{s,s'} e^{-i\epsilon_{s,s'}\tau/\hbar} \la e,n|\epsilon_{s,s'}\ra\la\epsilon_{s,s'}|g,0\ra\right|^2,
\end{eqnarray}
which may be analitically calculated to give
\begin{eqnarray}
\label{pe0}
P_{e0}&=&\left(\frac{\Omega_R}{2\Omega}\right)^2
\left[\left(\cos \frac{\nu_+\tau}{2}-\cos \frac{\nu_-\tau}{2}\right)^2\right.\nonumber\\
&+\!&\!\!\left.\left(\!\frac{\omega_T+\Omega}{\nu_+}\sin{\frac{\nu_+\tau}{2}}-
\frac{\omega_T-\Omega}{\nu_-}\sin{\frac{\nu_-\tau}{2}}\right)^2\right]\!\!,
% \\&\approx&\left(\frac{\Omega_R}{\Omega}\right)^2\sin^2\frac{(\nu_+-\nu_-)\tau}{4}
\\P_{e1}\!&=&\!\left(\frac{\eta\Omega_R}{2\Omega}\right)^{\!2}
\nonumber\\
&\times&\left(
\frac{\Omega-\Delta}{\nu_+}\sin{\frac{\nu_+\tau}{2}}+\frac{\Omega+\Delta}{\nu_-}\sin{\frac{\nu_-\tau}{2}}\right)^2. 
\label{pe1}
\end{eqnarray}
These are ``exact'' results within the LD and four-level approximations.
The oscillations in $P_{e0}$ and $P_{e1}$ may thus be viewed as interferences  
among the dressed states contributions and be characterized by 
frequencies $\nu_\pm$.  

Note also that the expressions (\ref{pe0}) and (\ref{pe1}) are valid for lasers of arbitrary intensity. 
In particular, transitions 
to higher order sidebands which in principle are off-resonant when $\Delta=0$, become important
when the ``Rabi Resonance'' condition $\Omega_R=\omega_T$ is fullfilled. In
this case $P_{e1}$ reduces to
\begin{equation}
P_{e1}\approx\frac{1}{4}\sin^2 \frac{\eta\Omega_R t}{2}.
\end{equation}
which shows that terms which are in principle off-resonant lead to resonant effects under certain conditions,
see also \cite{JPK00,MC99,APS03,LM06_a,LM06_b}.

Expressions (\ref{pe0}) and (\ref{pe1}) can be further simplified by performing an expansion in power series of the LD
parameter. To leading order in $\eta$, 
\begin{eqnarray}
P_{e0}&\approx&\left(\frac{\Omega_R}{\Omega}\right)^2\sin^2\frac{\Omega t}{2}
\left[1-\left(\frac{\Omega t}{2}\right)\frac{\eta^2\Omega_R^2}{\omega_T^2}
\cot\frac{\Omega t}{2}\right],\nonumber\\
P_{e1}&\approx&\left(\frac{\eta\Omega_R}{2\Omega}\right)^2
\left[A_+\sin \frac{(\omega_T+\Omega)t}{2}+ A_-\sin\frac{(\omega_T-\Omega)t}{2}\right]^2,\nonumber
\end{eqnarray}
with $A_{\pm}=\frac{\Omega\mp\Delta}{\omega_T\pm\Omega}$.
$P_{e1}(t)$ takes the form of a beating oscillation with a  
 fast frequency $\omega_T$ and a slow frequency $\Omega_R$.     

The expressions for the excited state probability simplify when the duration of the laser pulse is fixed.
If a $\pi$-pulse is applied ($\tau_\pi=\pi/\Omega_R$) we have that
\begin{eqnarray}
P_{e0}&\approx&\left(\frac{\Omega_R}{\Omega}\right)^2,\\
P_{e1}&\approx&\left(\frac{\eta\Omega_R}{2\Omega}\right)^2\left(A_+-A_-\right)^2\cos^2\frac{\omega_T\tau_\pi}{2},
\end{eqnarray}
which, near atomic resonance ($\Delta\sim 0$), can be written as
\begin{eqnarray}
P_{e0}&\approx&1-\frac{\Delta^2}{\Omega_R^2},\\
P_{e1}&\approx&\eta^2\left(\frac{\Omega_R^4}{\omega_T^4}+\frac{2\Omega_R^2\Delta}{\omega_T^3}
+\frac{\Delta^2}{\omega_T^2}\right)\cos^2\frac{\omega\tau_\pi}{2}. 
\end{eqnarray}
With these expressions for the excited state probabilities,
the shifted position of the central resonance follows from Eq. (\ref{derivative}): 
the central maximum in Fig. \ref{eprob_rabi} is pulled to the right, to higher frequencies,  by  
\begin{equation}
\delta(\tau_\pi)\approx
%\eta^2\frac{\Omega_R^4}{\omega_T^3}\cos^2\frac{\omega_T\tau}{2}\nonumber\\
%&\approx&
\Omega_R\eta^2\alpha^3\cos^2\frac{\omega_T\tau_\pi}{2},
\end{equation}
the same result obtained in the general $6$-state model calculation when a $\pi$-pulse
is applied, see Eq. (\ref{rabi_shift_pi_pulse}). 

%%%%%%%%%%%%%%%%%%%%%%%%%%%%%%%%%%%%%%%%%%%%%%%%%%%%%%%%%%%%%%%%%%%%%%%%%%%
\section{Derivation of the frequency shift in the Ramsey case}
\label{ramsey_shift_derivation}
From Eq. (\ref{ramsey_prob}) and with the (approximate) dressed energies (\ref{dressed_energies}) and
dressed states (\ref{dressed_states}) obtained in Appendix \ref{perturbation}, we may 
calculate the probabilities for the different $|e,n\ra$ states. To leading order in the LD
parameter and near atomic resonance $\Delta\sim 0$, they are given by
% %
\begin{eqnarray}
P_{e,n_0\pm1}&\approx&\frac{N\eta^2}{(1-\alpha^2)^2}
\left[\left(\alpha^2\cos\ttot+\alpha\sin\tfree\right)^2\right.\nonumber\\
&\pm& \frac{\alpha\Delta}{\omega_T}\left(2+T\Omega_R\right)\left(\cos\ttot+\alpha\sin\tfree\right)\nonumber\\
&\times&\left.\left(\alpha\cos\ttot+\sin\tfree\right)\right],
\nonumber\\
P_{e,n_0}&\approx&1-\left(\frac{1}{\Omega_R^2}+\frac{T}{\Omega_R}+\frac{T^2}{4}\right)\Delta^2,
\end{eqnarray}
with $N=n_0$ ($N=n_0+1$) for the red (blue) sideband. The presence of the blue and red sidebands will shift the 
position of the central resonance to a position satisfying the maximum condition (\ref{derivative}). This gives
a shift of
\begin{eqnarray}
\delta(T)&\approx&
% &&\left[\frac{\Omega_R}{\omega_T}\cos(\omega_T\tau+\frac{\omega_T T}{2})+\sin\frac{\omega_T T}{2}\right]\\&\approx&
% \eta^2\Omega_R^3\frac{\omega_T^2}{(\omega_T^2-\Omega_R^2)^2}\left(\frac{2}{2+\Omega_R T}\right)\times\nonumber\\
% &&\left(\cos{\frac{\omega_T T_t}{2}}+\frac{\Omega_R}{\omega_T}\sin\frac{\omega_T T}{2}\right)
% \left(\frac{\Omega_R}{\omega_T}\cos{\frac{\omega_T T_t}{2}}+\sin\frac{\omega_T T}{2}\right)\nonumber
\Omega_R\eta^2\frac{\alpha^2}{(1-\alpha^2)^2}\left(\frac{2}{2+T\Omega_R}\right)\nonumber\\
&\times&\left(\cos\ttot+\alpha\sin\tfree\right)\left(\alpha\cos\ttot+\sin\tfree\right).\nonumber
\end{eqnarray}
Keeping leading order terms in $\alpha$ if low intensity lasers are assumed ($\alpha=\Omega_R/\omega_T\ll1$) gives
the frequency shift 
\begin{eqnarray}
\delta(T)&\approx&
\Omega_R\eta^2\alpha^2\left(\frac{2}{2+\Omega_RT}\right)\left[\cos{\ttot}\sin{\tfree}\right.\nonumber\\
&+&\left.\alpha\left(\cos^2{\ttot}+\sin^2{\tfree}\right)\right],
\end{eqnarray}
which is Eq. (\ref{ramsey_shift}).
%

%%%%%%%%%%%%%%%%%%%%%%%%%%%%%%%%%%%%%%%%%%%%%%%%%%%%%%%%%%%%%%%%%%55


\newpage
\begin{thebibliography}{1}

\bibitem{LBMW03} D. Leibfried, R. Blatt, C. Monroe, and D. Wineland, 
Rev. Mod. Phys. {\bf 75}, 281 (2003). 

\bibitem{WMILKM98} D. J. Wineland, C. Monroe, W. M. Itano, D. Leibfried, B. E. King, and D. M. Mekhof,
J. Res. Natl. Inst. Stand. Technol. {\bf 103}, 259 (1998).

\bibitem{WI79} D. J. Wineland, W. M. Itano,
Phys. Rev. A {\bf 20}, 1521 (1975). 


\bibitem{D52} R. H. Dicke,
Phys. Rev. {\bf 89}, 472 (1952). 

\bibitem{madej} A.A. Madej and J.E. Bernard, ``Single Ion Optical Frequency Standards and
Measurement of their Absolute Optical Frequency'', in: Frequency Measurement
and Control : Advanced Techniques and Future Trends, Springer Topics in
Applied Physics , Andre N. Luiten editor, vol .79, (Springer Verlag, Berlin,
Heidelberg, 2001) p. 153-194.


\bibitem{HGRLBESB03} H. H\"affner {\it et al.},
Phys. Rev. Lett.  {\bf 90}, 143602 (2003). 

\bibitem{SHGRLEBB03} F. Schmidt-Kaler {\it et al.},
Europhys. Lett. {\bf 65}, 587 (2004).

\bibitem{EMSB03} J. Eschner {\it et al.}, 
J. Opt. Soc. Am. B {\bf 20}, 1003 (2003).

\bibitem{CBZ94} J. I. Cirac, R. Blatt, P. Zoller,
Phys. Rev. A {\bf 49}, R3174 (1994).

\bibitem{LM06_b} I. Lizuain, J. G. Muga, 
Phys. Rev. A {\bf 75}, 033613 (2007).


\bibitem{oskay06} W. H. Oskay  {\it et al.},
Phys. Rev. Lett. {\bf 97}, 020801 (2006).

\bibitem{diddams_science_01} S. A. Diddams {\it et al.},
Science {\bf 293}, 825 (2001). 

\bibitem{champenois04} C. Champenois {\it et al.},
Phys. Lett. A {\bf 331} 298-311 (2004).

\bibitem{riis04} E. Riis and A. G. Sinclair,
J. Phys. B: At. Mol. Opt. Phys {\bf 37} 4719-4732 (2004).

\bibitem{udem01} Th. Udem {\it et al.}
Phys. Rev. Lett.  {\bf 86} 4996 (2001).

\bibitem{LGRS04} V. Letchumanan, P. Gill, E. Riis, and A. G. Sinclair, 
Phys. Rev. A {\bf 70}, 033419 (2004). 

\bibitem{CZ95} J. I. Cirac and P. Zoller, 
Phys. Rev. Lett. {\bf 74}, 4091 (1995). 

\bibitem{kaler03} F. Schmidt-Kaler {\it et al.}
Nature {\bf 74}, Issue 6930, pp. 408-411 (2003) 

\bibitem{ramsey50} N. F. Ramsey, 
Phys. Rev. {\bf 78}, 695 (1950).

\bibitem{LM06_a} I. Lizuain, J. G. Muga, 
Phys. Rev. A {\bf 74}, 053608 (2006). 

\bibitem{JPK00} D. Jonathan, M. B. Plenio, and P. L. Knight, 
Phys. Rev. A {\bf 62}, 042307 (2000).

\bibitem{MC99} H. Moya-Cessa, A. Vidiella-Barranco, J. A. Roversi, 
D. S. Freitas, and S. M. Dutra, Phys. Rev. A {\bf 59}, 2518 (1999). 

\bibitem{APS03} P. Aniello, A. Porzio and S. Solimeno, 
J. Opt. B: Quantum Semiclass. Opt. {\bf 5}, S233 (2003).


% \bibitem{wineland84} D. J. Wineland,
% Science {\bf 226}, 395 (1984). 
% 
% \bibitem{G05} P. Gill,
% Metrologia {\bf 42}, S125-S137 (2005). 
% 
% \bibitem{margolis04} H. S. Margolis {\it et al.},
% Science {\bf 306}, 1355 (2004)






\end{thebibliography}
\end{document}